# Epitaxial graphene transistors on SiC substrates

Jakub Kedzierski, Pei-Lan Hsu, Paul Healey, Peter Wyatt, and Craig Keast
MIT Lincoln Laboratory, Lexington MA 02420
Mike Sprinkle, Claire Berger, and Walt de Heer
Georgia Institute of Technology, Atlanta GA 30332

**Abstract**:

This paper describes the behavior of top gated transistors fabricated using carbon, particularly epitaxial graphene on SiC, as the active material. In the past decade research has identified carbon-based electronics as a possible alternative to silicon-based electronics. This enthusiasm was spurred by high carbon nanotube carrier mobilities. However, nanotube production, placement, and control are all serious issues. Graphene, a thin sheet of graphitic carbon, can overcome some of these problems and therefore is a promising new electronic material.

Although graphene devices have been built before, in this work we provide the first demonstration and systematic evaluation of arrays of a large number of transistors entirely produced using standard micro-electronics methods. Graphene devices presented feature high-k dielectric, mobilities up to 5000 $cm^2$/Vs and, $I_{on}/I_{off}$ ratios of up to 7, and are methodically analyzed to provide insight into the substrate properties. Typical of graphene, these micron-scale devices have negligible band gaps and therefore large leakage currents.

# Epitaxial graphene transistors on SiC substrates

Jakub Kedzierski, Pei-Lan Hsu, Paul Healey, Peter Wyatt, and Craig Keast
MIT Lincoln Laboratory, Lexington MA 02420
Walt de Heer, and Claire Berger
Georgia Institute of Technology, Atlanta GA 30332

**Introduction**:

Carbon based transistors have attracted significant research interest due to their versatility and high intrinsic mobility. On the nano-scale, carbon can take many forms, such as conducting or semiconducting nanotubes, $C_{60}$ spheres, wide band-gap diamond, or graphite. Carrier mobility in graphitic forms of carbon, such as nanotubes and thin graphene sheets can be very high, on the order of 10,000 $cm^2/Vs$ [1,2]. This is at least 10 times higher than silicon [3]. Until recently a significant portion of research into carbon devices has focused on carbon nanotubes [4-6]. While carbon nanotubes are arguably the most versatile form of graphitic carbon, their versatility is not necessarily an advantage. It has become clear that controlling all the relevant characteristics in carbon nanotubes, i.e. their position, radius, chirality, length, doping, and strain, is a very difficult research problem.

An alternative form of carbon is graphene. Graphene is a horizontally extended single atomic layer of graphite. While graphene shares the high mobility of carbon nanotubes, it is planar and can be produced in extended sheets, making its integration into a future technology potentially easier [7,8]. The most serious problem of carbon nanotubes, control of position, radius, chirality, and length is largely eliminated, and it is much easier to imagine creating a substrate of graphene on which an electronic device technology can be based [7,9]. Graphene devices are, of course, not without their own challenges. Graphene is a semimetal with a zero band-gap. This makes it possible to control the conduction of graphene with the use of a gate, but makes it impossible to turn the conduction 'off' below a certain limit [8]. The difficulty of overcoming this fundamental challenge should not be understated, however several methods of producing a band-gap in graphene have already been proposed and demonstrated [10-12]. Even if inducing a sufficiently large band-gap in graphene is not possible, its high mobility may make it an excellent material for RF electronic devices, in which relatively large leakage currents can be tolerated.

Recently graphene devices have been built on thin exfoliated sheets of highly oriented pyrolytic graphite (HOPG) [1,13]. While this is an excellent method for obtaining single devices for electrical analysis, building an entire electronic technology on such a method is problematic, because large scale uniformity of exfoliated flakes is poor. It is more desirable to prepare an entire substrate of graphene on an insulating substrate, then pattern the graphene in areas where it is required, with a process flow similar to one used for Si devices on silicon-on-insulator substrates. An entire substrate of graphene on insulator makes it possible to integrate large scale circuits, not just individual devices. A process for generating graphene on insulator, or more specifically multilayer epitaxial graphene (MEG) on SiC (MEG/SiC), has been developed through the high temperature sublimation of silicon from SiC [14].

In this paper the MEG/SiC substrates, fabricated at Georgia Tech, are used to develop a device technology capable of integrating hundreds of graphene devices over many square mm. Graphene transistors presented feature high-k dielectric, mobilities up to 5000 $cm^2/Vs$, and $I_{on}/I_{off}$ ratios of up to 7, and are methodically analyzed to provide insight into the substrate properties.

**Materials and Substrate Preparation**

Graphitic films on SiC substrates were prepared by solid-state decomposition of single crystal 4H-SiC (0001) in vacuum. The method involves an inductively heated vacuum furnace in which 3.5 mm X 4.5 mm X 0.3 mm SiC chips, are heated to about 1400 °C. In this process, Si sublimes to produce carbon-rich surfaces that subsequently graphitize. The graphitization produces epitaxially ordered stacked layers of graphene, with a high structural coherence length [15]. Figure 1 shows this multi-layered epitaxial graphene (MEG) at the different stages of preparation at Georgia Tech. Once complete the substrate chips were transferred to MIT Lincoln Laboratory for device integration.

Prior to integration G/SiC chips were characterized using optical and AFM measurements at MIT LL. It is important to note that SiC is not symmetric, the Si – C bond in the [0001] direction has an asymmetry just due to the fact that one end is Si and the other is C. Consider cleaving the SiC lattice by breaking that particular bond along the (0001) plane. This cleave results in two interfaces, the silicon terminated surface is called the Si-face, and the carbon terminated surface is called the C-face, figure 2. A typical SiC wafer will have a Si-face in the front with a [0001] normal, and a C-face in the back with a [000-1] normal. During silicon sublimation graphene layers are generated on both faces of the SiC wafer, however the film generated on the C-face has different properties from the film generated on the Si-face.

Results from three chips will be described in this paper, C712, C781, and S767. The C in the first 2 chips indicates that the C-face graphene was used for device integration, while the S in the third sample indicates that the Si-face was used. Figure 3 shows the optical and AFM data for sample C781 and S767. As typical for the C-face graphene, C781 shows significant optical brightness variation. This brightness variation is caused in large part by the local thickness variations in the graphene layer, as shown in figure 4. Figure 4 indicates a correlation between brightness and measured MEG thickness, for both of the C-face samples. The normalized graphene brightness is defined as the difference in the brightness between the graphene and the adjacent SiC, divided by the SiC brightness. The MEG film thickness was measured by locally etching the graphene with an $O_2$ plasma and measuring the resulting edge height with an AFM. On C781 the MEG film thickness varied from about 3 nm to about 10 nm, while on C712 it varied from about 3 nm to about 15 nm. The characteristic length scale of the variation was approximately 10μm.

The AFM scans of C781 indicate a microstructure consisting of large domains. The exact microstructure can not be ascertained from AFM alone. When interpreting AFM data it is important to remember that steps in the top of the MEG film maybe the result of the SiC steps or structure that is buried beneath. The exact microstructure of the MEG grains, including the manner in which they are interconnected, requires further investigation.

The S767 Si-face looks significantly different from the C-face samples, both optically and in the AFM. As is typical for furnace grown Si-face samples, the steps are more regular and uniform, having a consistent height and direction. Note that the height of each step is many lattice units of SiC, indicating that the atomic steps have coalesced during the preparation anneals. In addition to the steps, the Si-face has a complex structure of plateaus that populate each terrace. Optically the Si-face looks more uniform in brightness and is darker. Correspondingly the AFM edge height measurements indicate that indeed the MEG film is significantly more uniform and thinner, varying from about 1.6 nm to about 2.2 nm. The exact microstructure of the graphene on the Si-face also requires further investigation. However it is known from resistivity and STM analysis that the graphene layers are largely continuous over the edges of the SiC on both faces [7,16].

**Device integration**

After characterization, G/SiC chips were mounted on 150-mm silicon carrier wafers using epoxy bonding. This was done so that the silicon fabrication tools are able to process the small chips. First, alignment marks were defined with standard g-line lithography and etched into the G/SiC with a $Cl_2$/He plasma etch. These marks are required because the active MEG layer is too difficult to see optically for consistent alignment of subsequent layers. Following the alignment mark etch, the resist was stripped in 80°C sulfuric acid; this strip did not affect the appearance or resistivity of the MEG layer. Next, the active MEG layer was patterned using a low energy $O_2$ plasma etch. The source/drain layers were deposited directly on the MEG film layer and consisted of 2 nm Ti and 20 nm of Pt, defined using a lift off process. A 40 nm $HfO_2$ layer was then deposited over the entire chip, using thermal evaporation. The $HfO_2$ film was verified to have a dielectric constant of 23 via a capacitive measurement of a finished device. Finally, a 100 nm Al gate was deposited and defined using lift-off. The AFM of a finished device is shown in Figure 5.

The mask pattern used in this experiment contained approximately 100 devices, with different gate lengths, graphene widths, and alignment conditions. The nominal device was a one with a source to drain spacing of 10 μm, a graphene width of 5 μm, and a 15 μm gate overlapping the source and drain by 2.5 μm on each face. Hundreds of transistors where fabricated, with functional yield as high as 95% for some samples.

**Results and Discussion (Carbon face)**

The $I_d$-$V_g$ of a representative C-face MEG transistor is shown in Figure 6. Graphene devices generally exhibit a V shaped $I_d$-$V_g$ characteristic [1,13]. This is consistent with the semimetal nature of graphene and the ambipolar nature of the Pt contacts. As a semimetal graphene has no bandgap, thus the source/drain can be a source of either electrons or holes. As such the terminology of source and drain, adopted from semiconductor devices, is not completely appropriate, since in a single device both sides act as the source and the drain, just for carriers of different polarities. In such ambipolar devices it is more appropriate to call the low voltage contact an electron-source and the high voltage contact the hole-source.

As expected, our MEG devices exhibit increasing conduction for both positive and negative gate voltages. For negative gate voltages the conduction is dominated by holes. An increasingly negative gate voltage causes a corresponding increase in the accumulated hole concentration, thus leading to more conduction. Similarly for positive gate voltages conduction is dominated by electrons, with higher positive values leading to an increase in the accumulated electron concentration. At a certain gate voltage the conduction has a minimum. This occurs roughly where the conduction contributions of the holes and electrons are the same.

Figure 7 shows the source to drain conductivity as a function of gate voltage for a collection of identically fabricated devices from sample C712 and C781. It is clear that from device to device and from sample to sample the electrical characteristics are significantly different. On sample C712 minimum conduction, $\sigma_{dmin}$, varies from 400 μS to 7 mS, while the mobilities range from 500 $cm^2/Vs$ up to 5000 $cm^2/Vs$. On C781 $\sigma_{dmin}$ varies from 700 μS to 2.5 mS, while the mobilities range from 700 $cm^2/Vs$ to 3000 $cm^2/Vs$. The mobility is calculated as $(d\sigma_d/dV_g)(WL/C)$, with the derivative calculated from the steepest 3 adjacent data points for each branch. W is the MEG width, L is the length between the Pt source and the Pt drain, and C is the MEG to gate capacitance. The conductivity, $\sigma_d$, is calculated from the drain current, drain voltage, and device geometry, with a correction for the measured Pt series resistance. All conductivities are given per square of MEG film active area, as is the convention.

Since one of the most important properties of any electronic technology is the ability to make devices with similar properties, understanding what causes this variation between nominally identical devices is critical. Figure 8 shows the mobility of the electron and hole branches as a function of $\sigma_{dmin}$. There is no significant correlation between these two parameters. On average, mobility of devices on C712 is higher than on C781, as is $\sigma_{dmin}$. Inside each sample the authors looked for measured parameters that would correlate to mobility. No parameter correlated well, but a marginal correlation was found to the MEG film thickness uniformity, Figure 9. Here uniformity is defined as the inverse of the full width at half max of the MEG film thickness histogram.

Although factors influencing mobility still need to be investigated, the initial data shown here and in other experiments is very promising. Carrier mobilities for gate induced charges of up to 5000 $cm^2/Vs$, with a poor dielectric, are at least an order of magnitude higher than what would be expected in bulk silicon under similar circumstances. It is also remarkable that these room-temperature accumulation-layer mobilities, for processed samples, are comparable to Hall mobility values measured at low temperature in similar multilayered graphene films [2], and exfoliated single sheet graphene [8,17].

The second important device parameter is the conduction minimum, $\sigma_{dmin}$. $\sigma_{dmin}$ determines how well a graphene device can turn off. On the C-face $\sigma_{dmin}$ varies more than an order of magnitude between different devices. Unlike mobility the $\sigma_{dmin}$ correlates well to a measured device parameter: the MEG film thickness. Figure 10 shows this correlation between $\sigma_{dmin}$ and thickness for several of the more uniform devices on C781. The MEG film thickness used here is obtained from AFM edge step measurements averaged over the entire active graphene edge, the error bars indicate uncertainty. As seen in figure 10 the minimum conduction increases by roughly 150 μS per graphene monolayer. This is consistent with a few $e^2/h$ of conduction per sheet found in other experiments [1,10-11]. A similar relation can be found for devices on sample C712, with 270 μS of conduction per graphene monolayer. The different conduction per monolayer between the samples indicates that graphene microstructure plays an important role for $\sigma_{dmin}$, as well as for mobility.

**Results and Discussion (Silicon Face)**

The transistors fabricated on the silicon face of the MEG/SiC substrate are significantly different from the carbon face devices. Figure 11 shows the $I_d$-$V_g$ of a representative Si-face MEG device. Si-face devices show lower mobility, have significantly lower $\sigma_{dmin}$, and higher On/Off current ratio than C-face devices. In addition the electron conductivity saturates with increasing gate voltage, in a behavior that was also observed for C-face devices, but to a lesser extent.

A collection of 'identical' devices is shown in figure 12. Si-face devices show much more consistent behavior, with mobilities and $\sigma_{dmin}$ varying only by a factor of 2. Mobilities on the Si-face vary from 600 $cm^2/Vs$ to 1200 $cm^2/Vs$, while the $\sigma_{dmin}$ varies from 125 μS to 250 μS. The lower variation of these parameters reflects the fact that Si-face MEG film is more uniform. The thickness uniformity is better, varying only from about 1.5 nm to 2.5 nm, and the SiC step direction and spacing is more regular. The plot of mobilities vs. $\sigma_{dmin}$ is shown in figure 13. The data is much more clustered than on the C-face but a correlation is still not apparent. What is apparent is that despite the graphene being smoother on the Si-face the mobility is significantly lower than on the C-face. The reasons for this are still under investigation, but are likely due to the differences between Si-face and C-face graphene microstructure [15].

The $\sigma_{dmin}$ values, when correlated to MEG thickness, correspond to approximately 50 μS per monolayer of graphene. This is a significantly lower level of conduction per layer than for a C-face monolayer, suggesting that the intrinsic carrier mobility is significantly lower on the Si-face, just like the accumulated carrier mobility. Not unexpectedly, the conduction per sheet for the different samples C712, C781, and S767 (270, 150, and 50 uS/monolayer) varies approximately in the same ratio as the average accumulation mobility for each sample (2800, 1800, 800 $cm^2/Vs$).

The nominal devices on S767 were fabricated in two configurations, one with the current flow approximately aligned parallel to the substrate steps, and another with the current flow approximately perpendicular to them. If the microstructure of graphene had breaks at the step edge one would expect that the carrier mobility would be significantly lower for the perpendicularly aligned devices. This was not the case. While the average perpendicular device mobility was slightly lower than the parallel device mobility the difference was not statistically significant. This indicates that graphene film is largely continuous over SiC steps, in agreement with previous work [7,16].

**Technology discussion**

Although promising, graphene based electronics faces many obstacles before it can become a competitive technology. Minimum conduction has to be decreased, device to device variability has to be controlled, and a stable gate dielectric must be found. However the chip level integration of hundreds of graphene devices on insulating SiC substrates is a step towards making graphene technology possible. The main driver for a graphene technology is clearly mobility. Even in this preliminary experiment mobilities up to 5000 $cm^2/Vs$ have been achieved. This is already 10 times better than silicon technology which has had decades of optimization. It doesn't seem unreasonable to expect that after thorough investigation and process optimization, graphene devices will have mobilities over 10,000 $cm^2/Vs$ [2,9]. The greatest obstacle to a graphene technology is the lack of a band-gap, and thus an inability to turn off conduction below a certain level. Although this is a fundamental issue with graphene, some solutions exist.

Methods of inducing a bandgap with nanopatterning, a vertical electric field, or uniaxial stress, have all been proposed or demonstrated [10-12]. It is likely that some method of obtaining an on/off ratio for current in the hundreds will be demonstrated in the near future.

## Conclusion

This paper presents one of the first comprehensive investigations on how an integrated graphene technology can be implemented on a SiC substrate. It is the first time, to the knowledge of the authors, that hundreds of carbon based devices have been successfully fabricated in an integrated technology. Device behavior has been described for multi-layered epitaxial graphene devices built on both the carbon and silicon faces of SiC. Trends in mobility indicate that the graphene microstructure is largely responsible for mobility variation between chips and faces, with the Si-face multi-layered graphene films having a significantly lower mobility than the C-face multi-layered graphene films. Film thickness uniformity also has an influence on mobility, with more uniform regions having higher mobility. Mobilities as high as 5000 $cm^2/Vs$ were achieved for some C-face devices, while Si-face devices demonstrated a consistent $I_{on}/I_{off}$ ratio of about 5. Minimum conduction between the source and drain was found to be strongly correlated to graphene film thickness.

It is the opinion of the authors that graphene holds great promise for future electronic technology. It has excellent thin film properties, films that are as thin as 0.4 nm have been shown to have high mobility [8,17]. This is in contrast to silicon where mobility rapidly degrades as a function of thickness at the nanometer scale. Graphene also has been show to be compatible with high-k dielectrics, thus gate dielectric scaling beyond the limits of $SiO_2$ is possible. This is again in contrast to silicon technology where the use of high-k dielectrics reduces mobility. Thus while it is too early to speculate on graphene replacing silicon as the material of choice for electronics, the potential of carbon based devices should not be underestimated.

## Acknowledgments

The MIT Lincoln Laboratory part of this work was sponsored by the Defense Advanced Research Projects Agency under Air Force contract number: FA8721-05-C0002. Opinions, interpretations, conclusions, and recommendations are those of the authors and are not necessarily endorsed by the United States

Government.  Thanks to members of the Microelectronics Laboratory at MIT Lincoln Lab for their crucial assistance in fabrication of these unique devices. The Georgia Tech team gratefully acknowledges financial support from NSF (NIRT grant No. 0404084, and MRI grant No. 0521041), Intel, the SRC-INDEX program and CNRS.

Figure 1

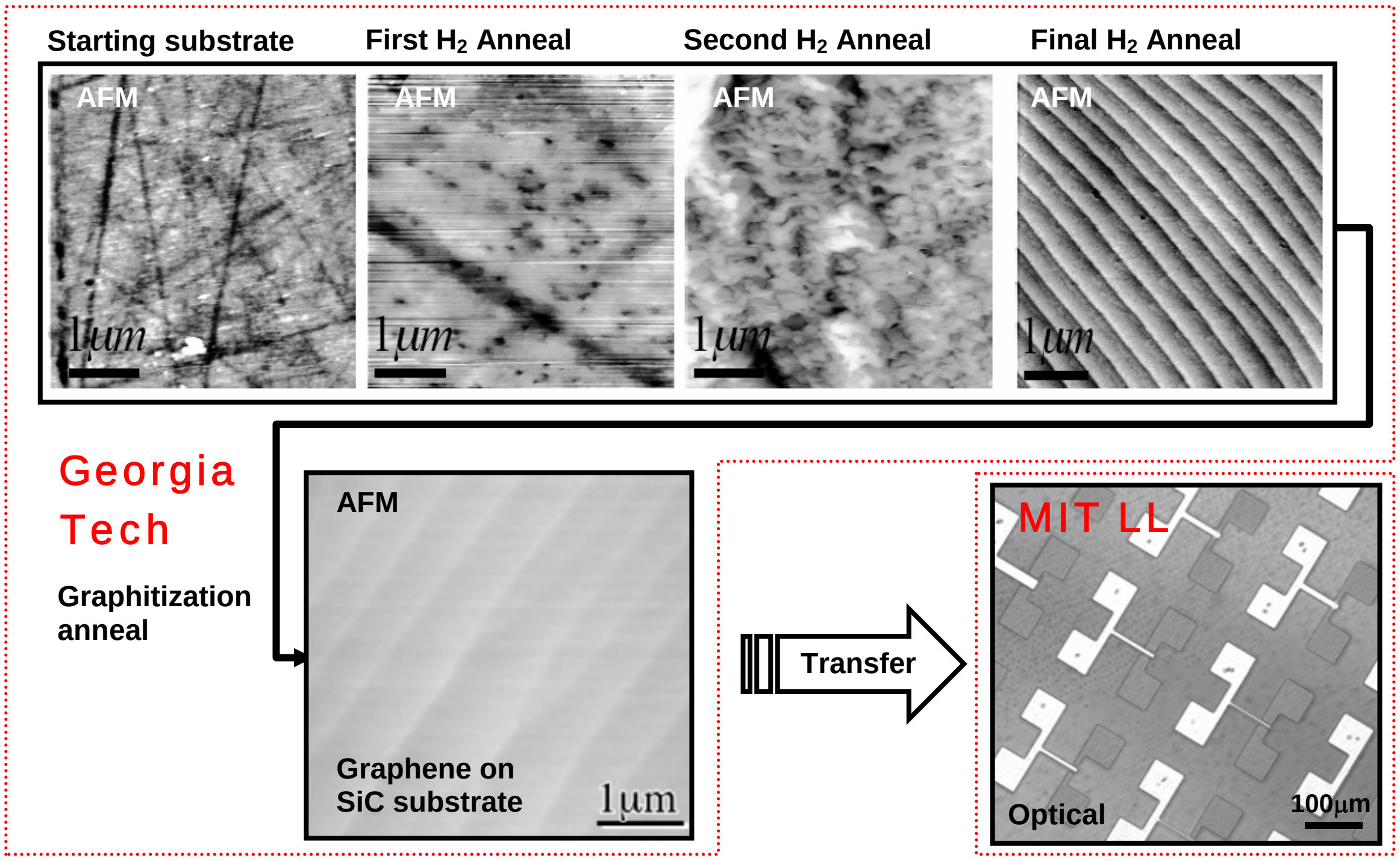

Figure 1: Preparation stages of the G/SiC substrate chips (4.5x3.5mm). A polished 4H-SiC substrate is annealed in subsequent $H_2$ anneals, until an aligned series of steps is formed. These steps are formed due to the agglomeration of atomic terraces caused by the miscut of the wafer surface from the (0001) plane. A graphitization anneal is then performed at around 1400°C to sublime the Si and form the graphene layers on SiC. The substrates are then transferred from Georgia Tech to MIT Lincoln Lab for integration.

Figure 2:

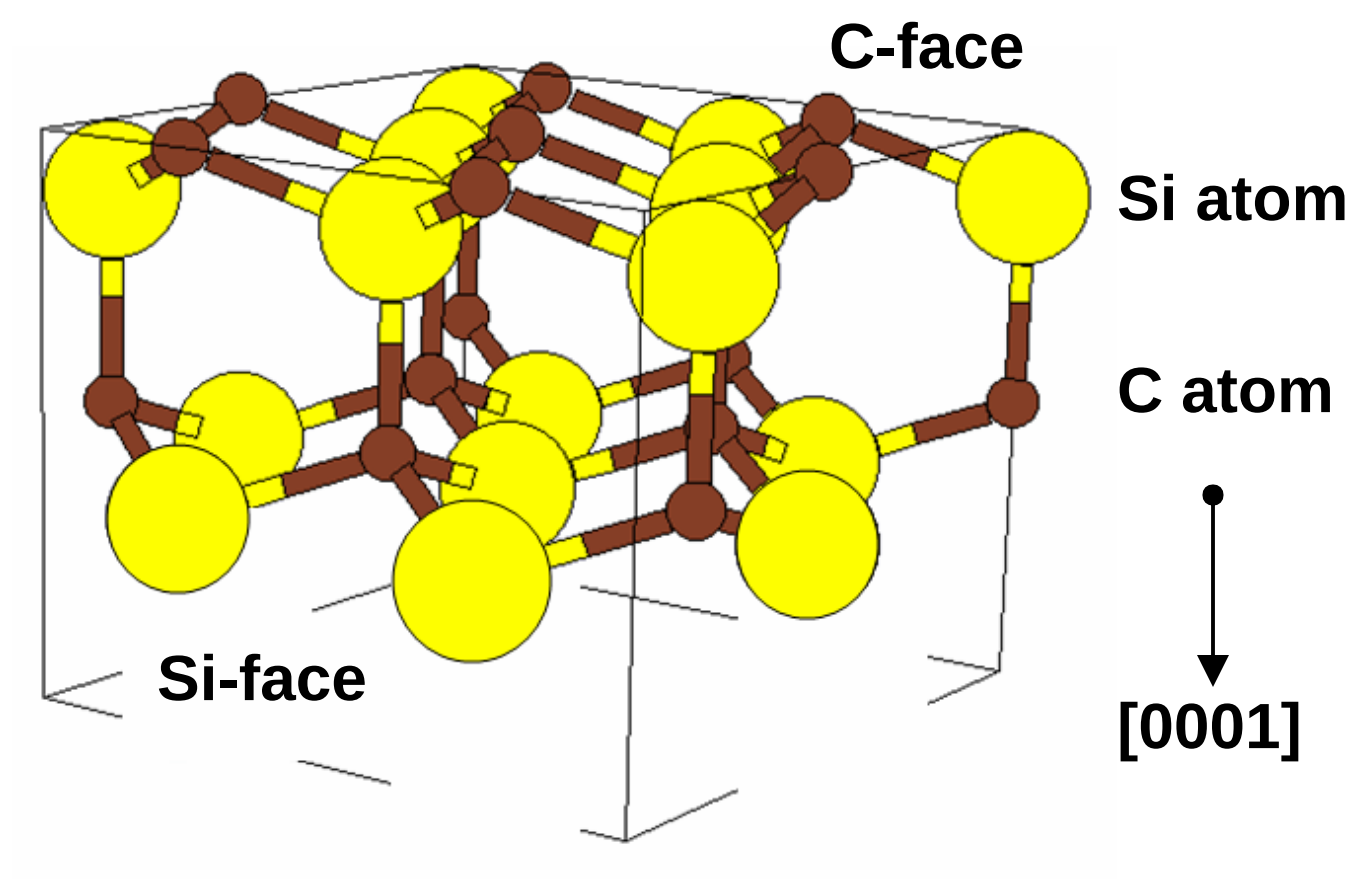

Figure 2: Crystal structure of SiC showing the two faces of the crystal cut along the (0001) plane. The [0001] directed face, terminated by Si atoms, is referred to as the Si-face, and the opposite face, terminated by C atoms, is referred to as the C-face.

Figure 3:

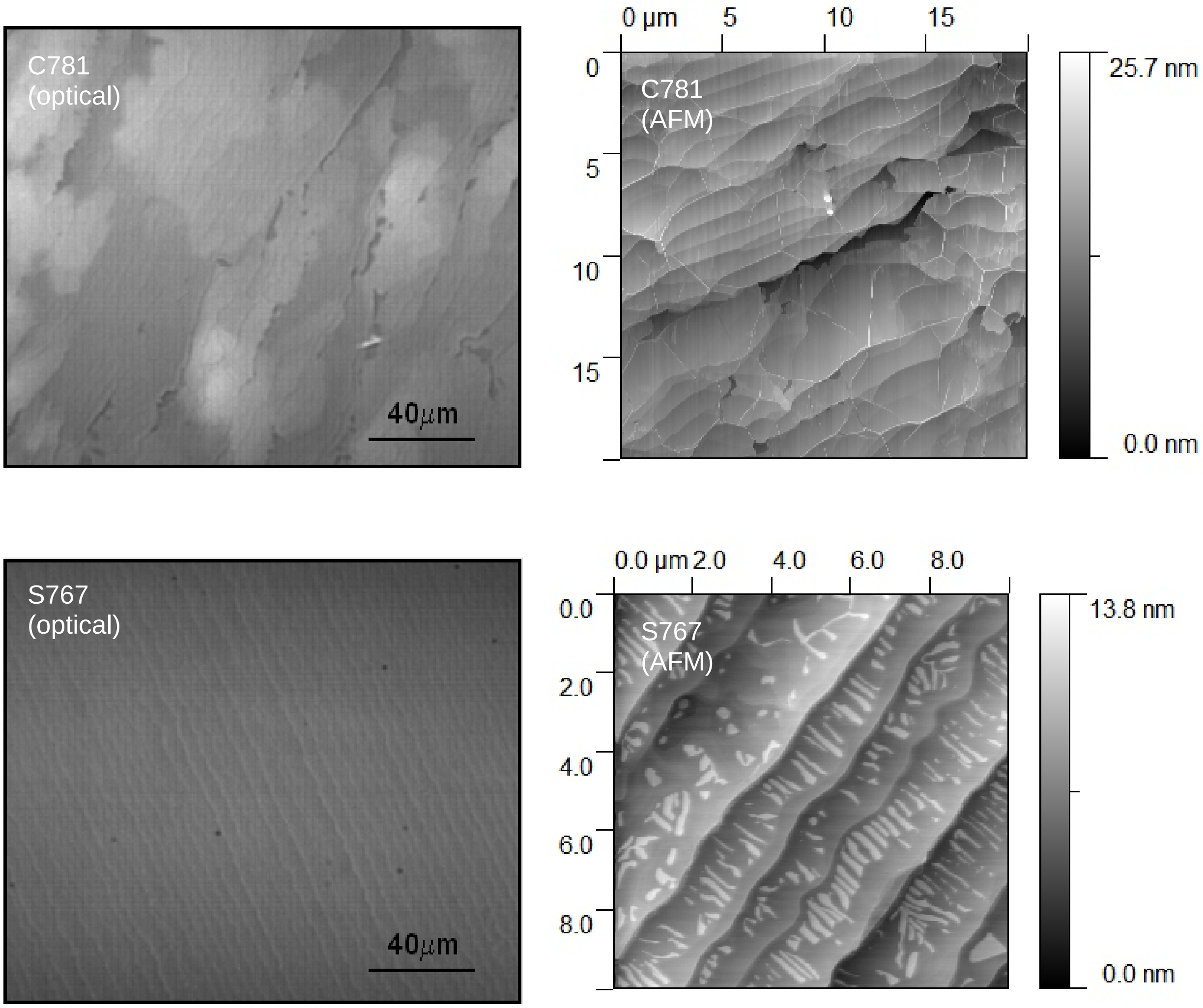

Figure 3: Optical and AFM micrographs of the G/SiC substrates. The C-face multi-layered graphene is shown for C781 and the Si-face multilayered graphene is shown for S767. These are the sides of the substrate that were processed in this experiment. C781 exhibits a flake-like structure, S767 is characterized by series of terraces. Both of these samples are typical of their respective face. It is important to note that the surface height variation shown in the AFM plots (fig 2) is not the graphene thickness variation (fig 3), but the total surface roughness, which includes multi-layered graphene thickness and significant roughness in the SiC underneath.

Figure 4:

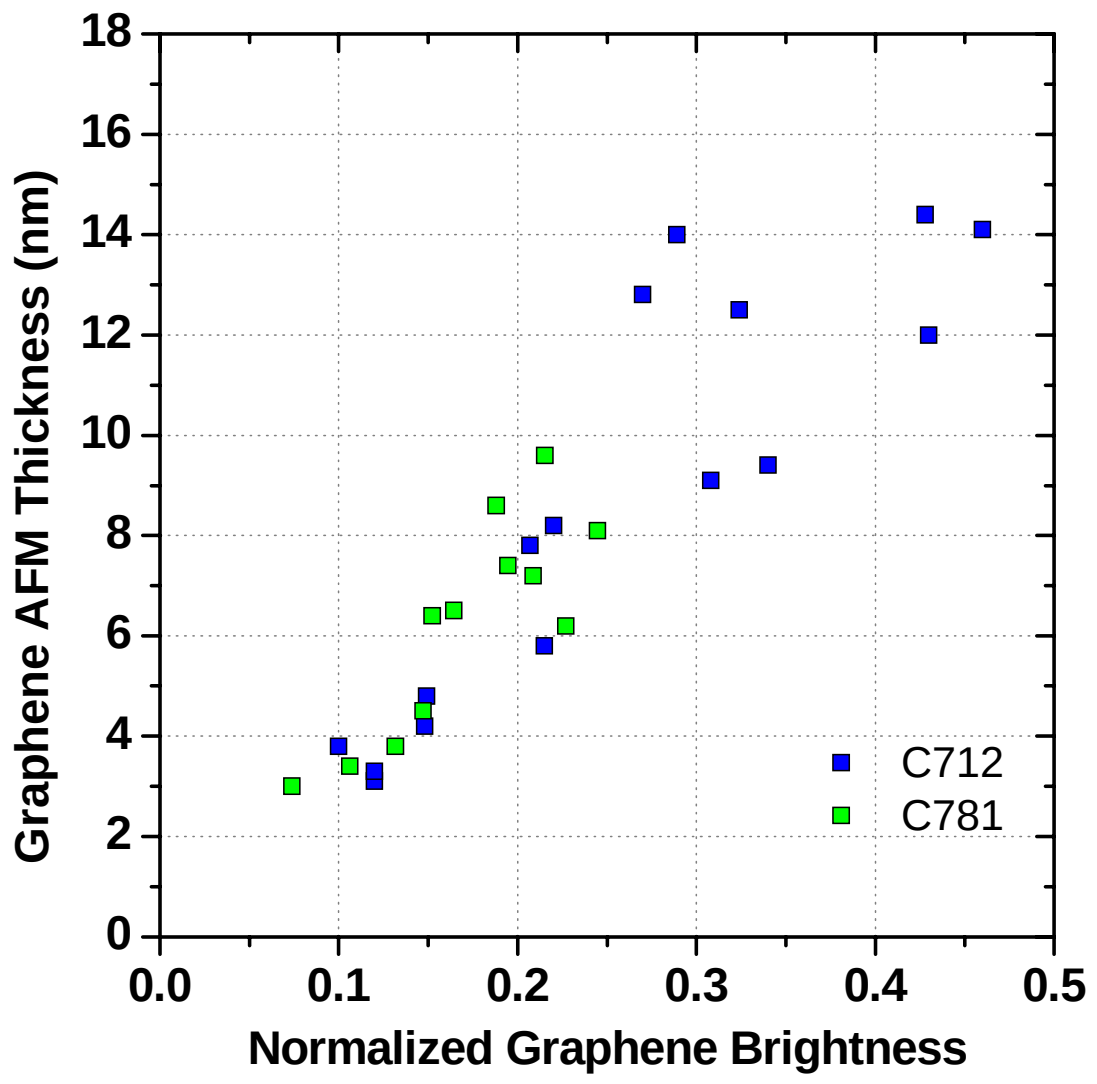

Figure 4: MEG film thickness as measured by AFM etched edge step heights, plotted vs. local normalized graphene film brightness. Normalized brightness is calculated as the MEG brightness minus the SiC brightness divided by the SiC brightness. The strong correlation shows that MEG film brightness is primarily determined by thickness.

Figure 5:

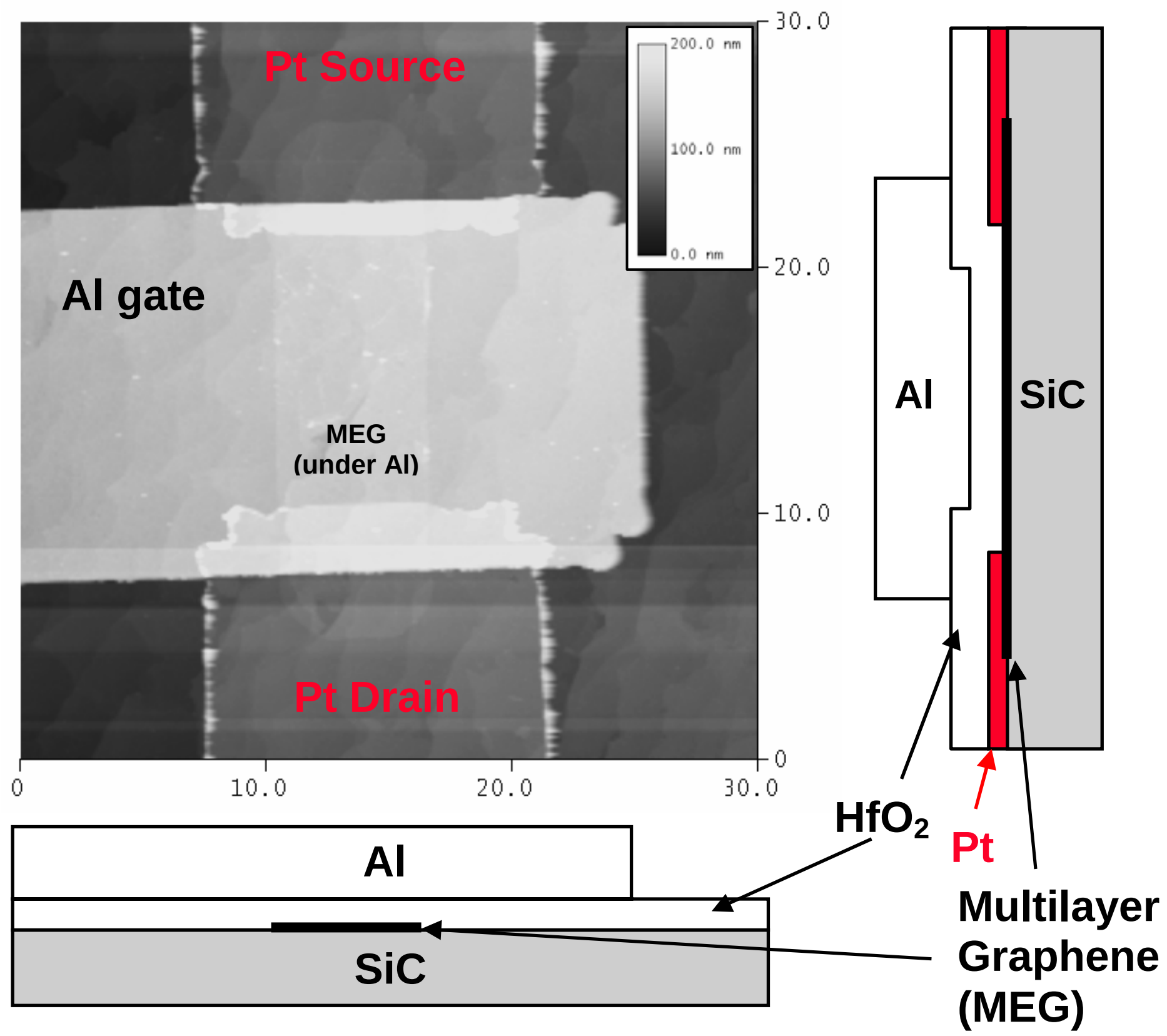

Figure 5: AFM scan of a finished nominal device on sample C781. The MEG film in this device is 7 nm thick. Schematic cross sections are shown to the bottom and right for clarity. Device electrical length, the space between source and drain is 12 um, and the MEG film width is 6 um.

Figure 6:

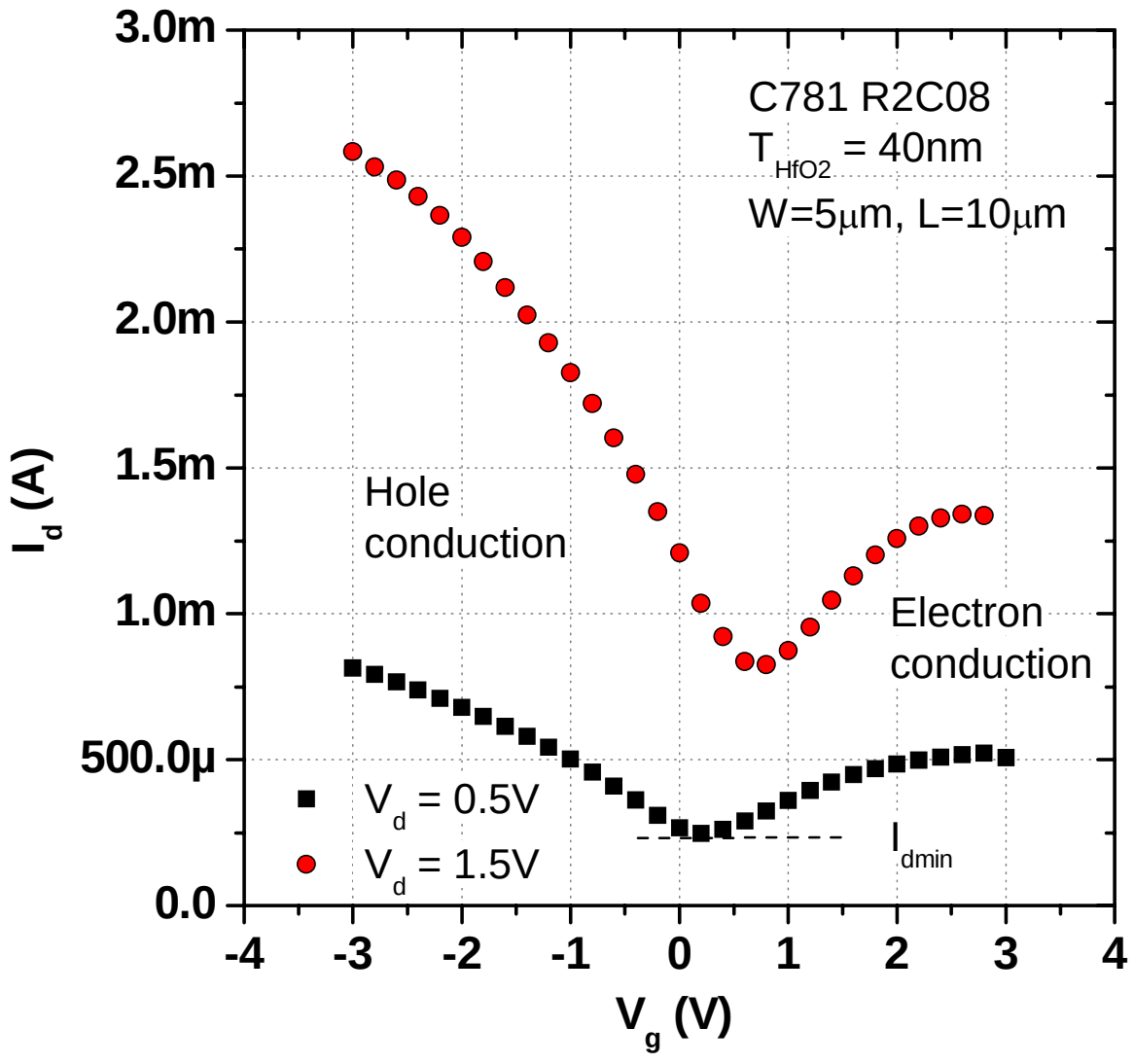

Figure 6: $I_d$-$V_g$ characteristics for a typical C-face MEG transistor, for two drain voltages. Hole concentration and conduction increase for negative voltages, electron concentration and conduction increase for positive voltages. A minimum conductivity is achieved when the two conductivities are approximately equal.

Figure 7:

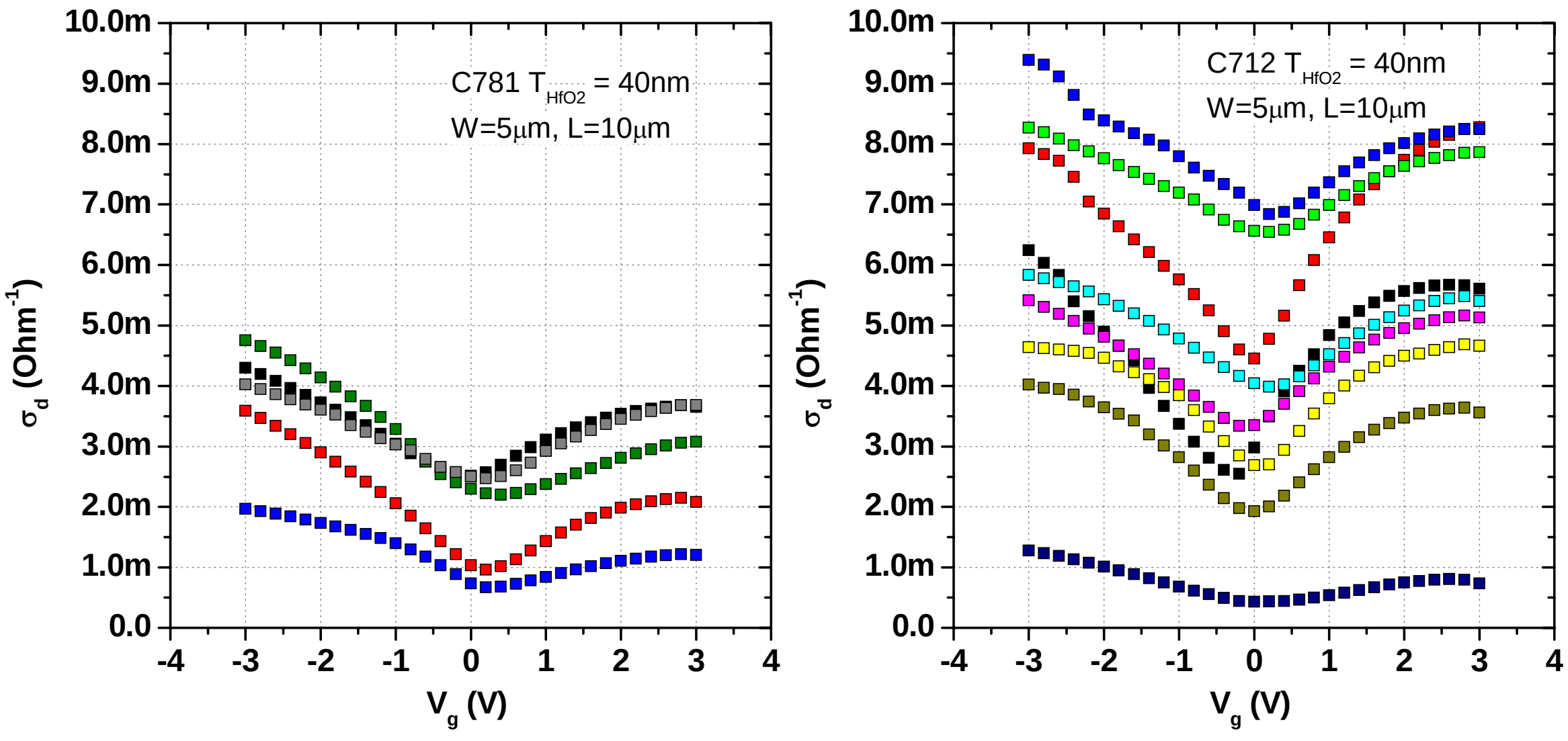

Figure 7: Conductivity characteristics for a set of identically processed devices with same electrical length and width, for two different C-face samples. On average, C781 had lower minimum conductance and slightly lower mobility than C712.  Conductivity units are milliSiemens per square of graphene.

Figure 8:

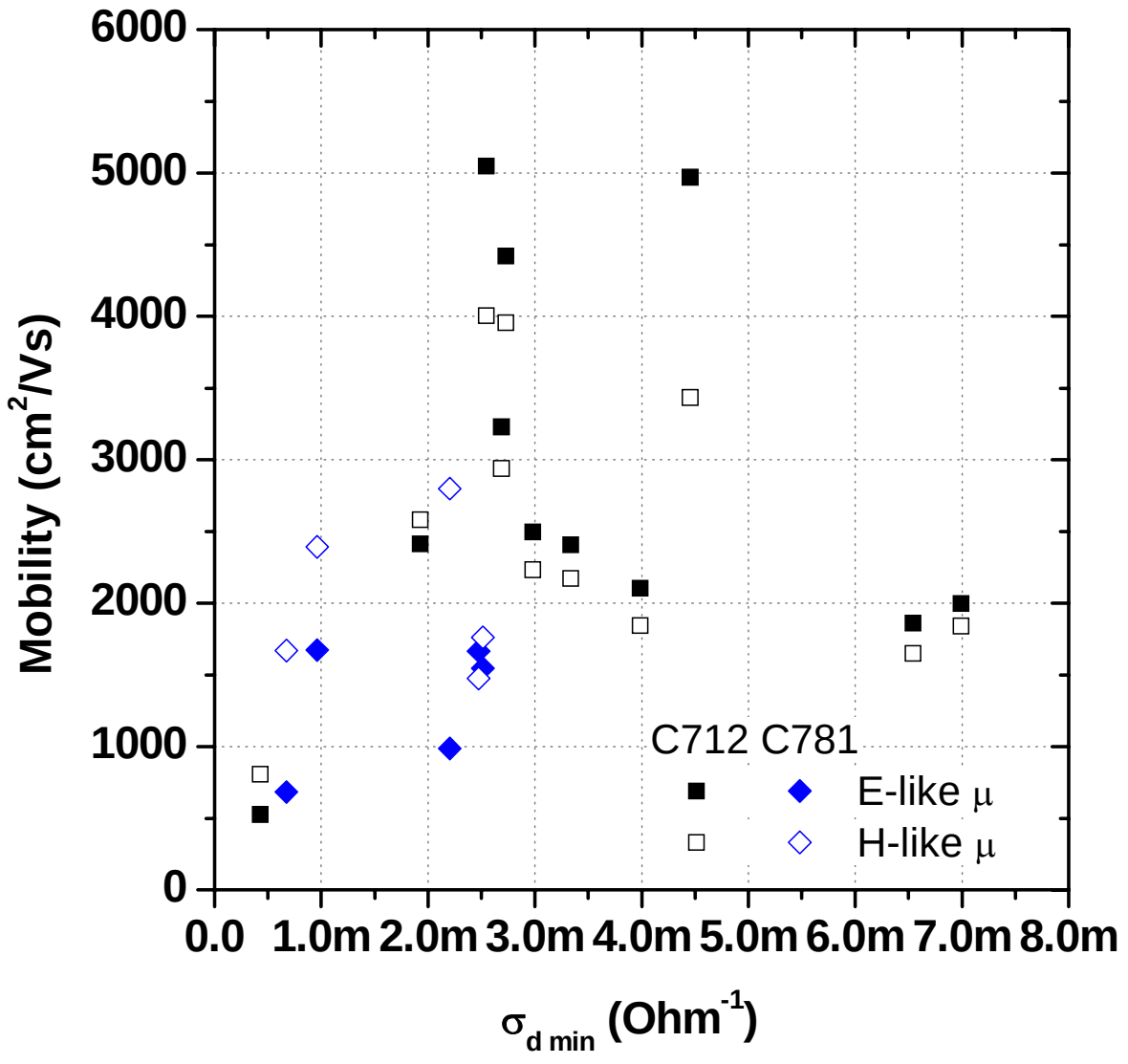

Figure 8: Carrier mobility for C-face samples plotted vs. minimum conductivity. No inter-sample correlation is found, although C781 has lower minimum conductivity and mobility on average, than sample C712. Mobility as high as 5000 $cm^2/Vs$ is achieved for some devices.

Figure 9:

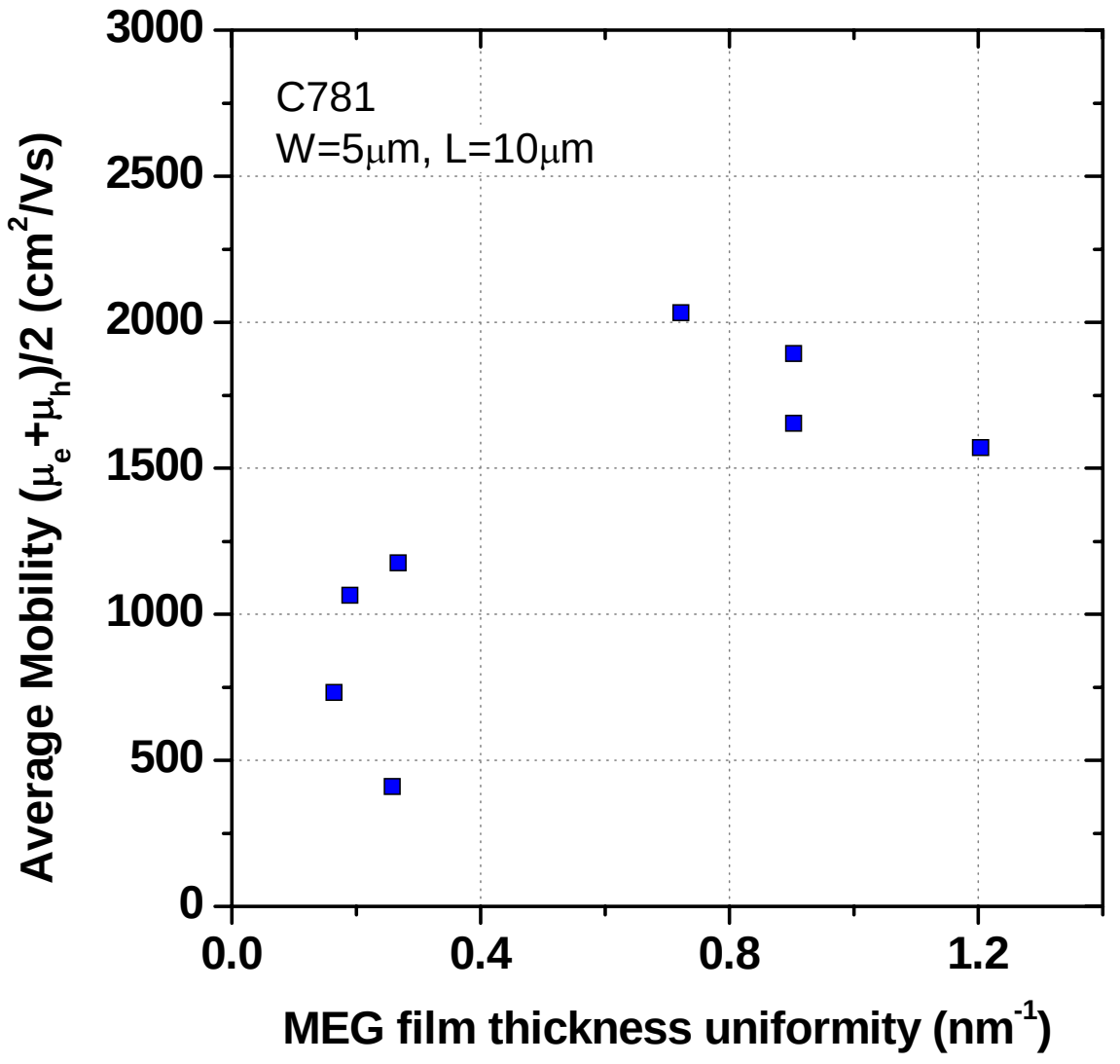

Figure 9: Average of hole and electron mobilities plotted vs. MEG film thickness uniformity. A weak correlation exists, with rougher samples displaying lower mobility. MEG film thickness uniformity is calculated as the inverse of the width at half max of the MEG thickness histogram for the device under measurement.

Figure 10:

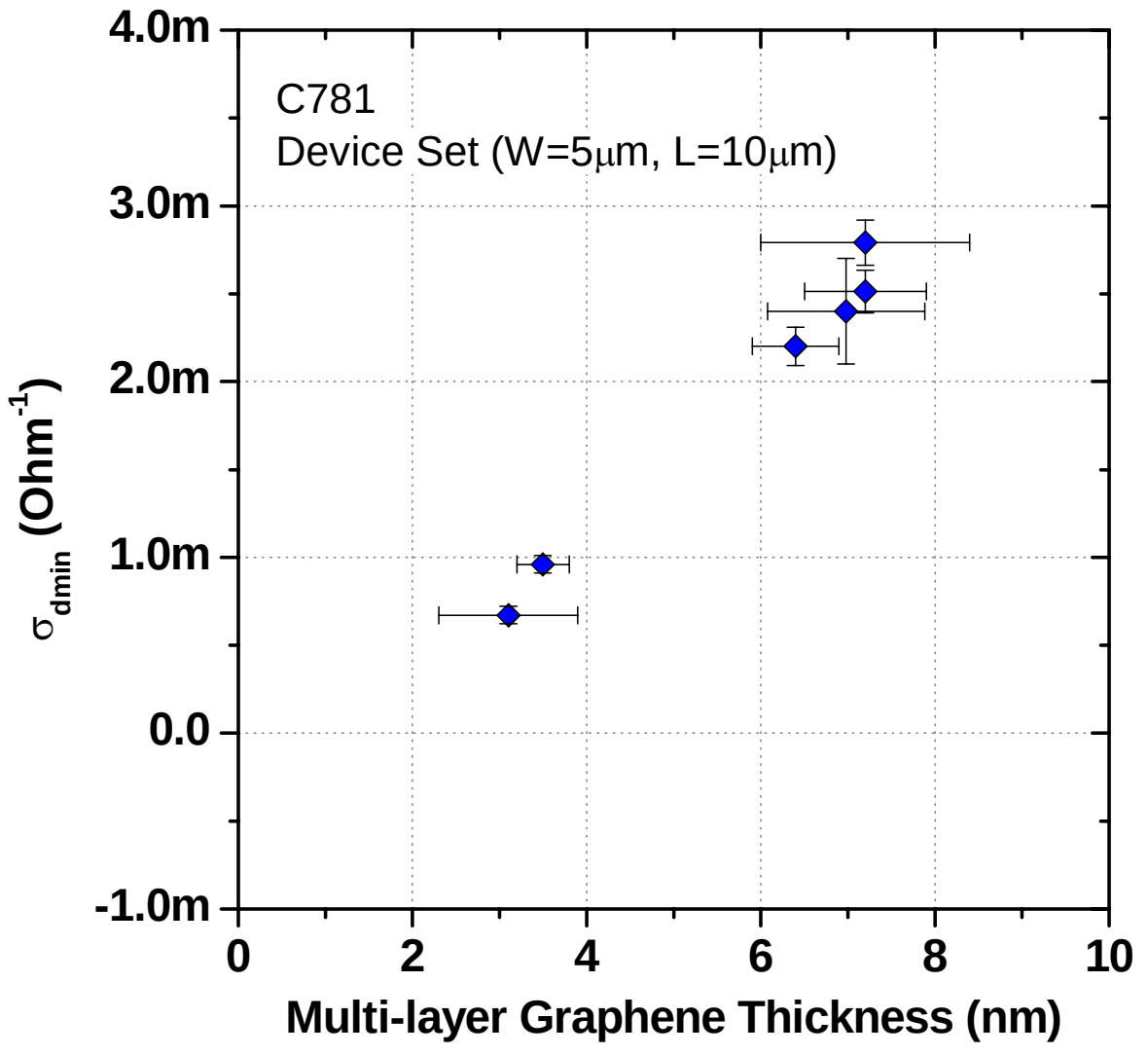

Figure 10: Minimum conduction vs. MEG film thickness on C781. Thickness is obtained from AFM step height measurements at the edge of the active film. Error bars indicate estimate of the error range given the variation in thickness across the device. A correlation is found between MEG thickness and conduction. In addition to the indicated measurement error, a small systematic error in the absolute value of the thickness may exist due to plasma processing at the edge of the MEG film.

Figure 11:

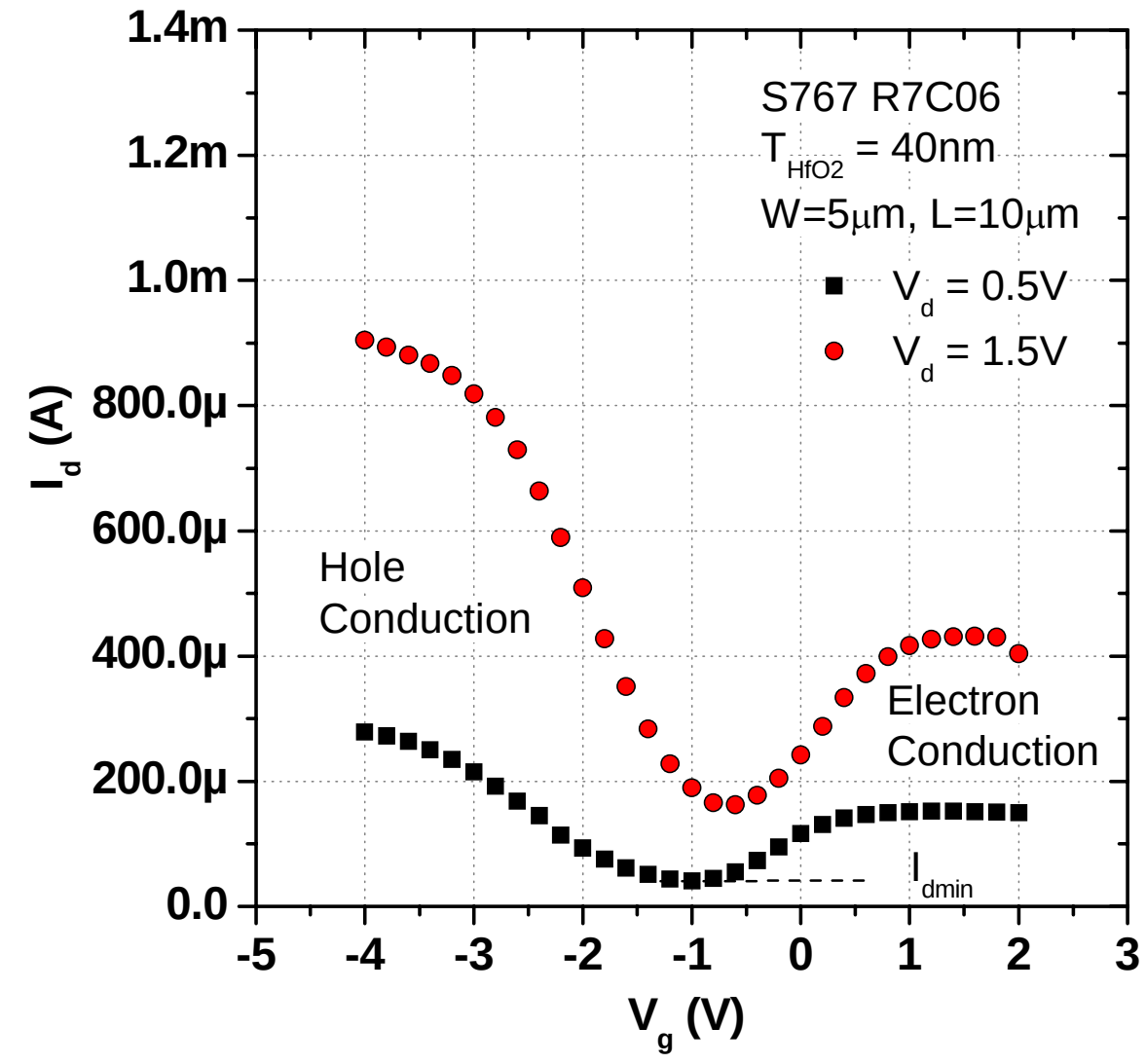

Figure 11: $I_d$-$V_g$ characteristics for a typical Si-face MEG transistor, for two drain voltages. Both electron and hole mobilities appear to decrease with increasing carrier concentration, much more so than for C-face devices.

Figure 12:

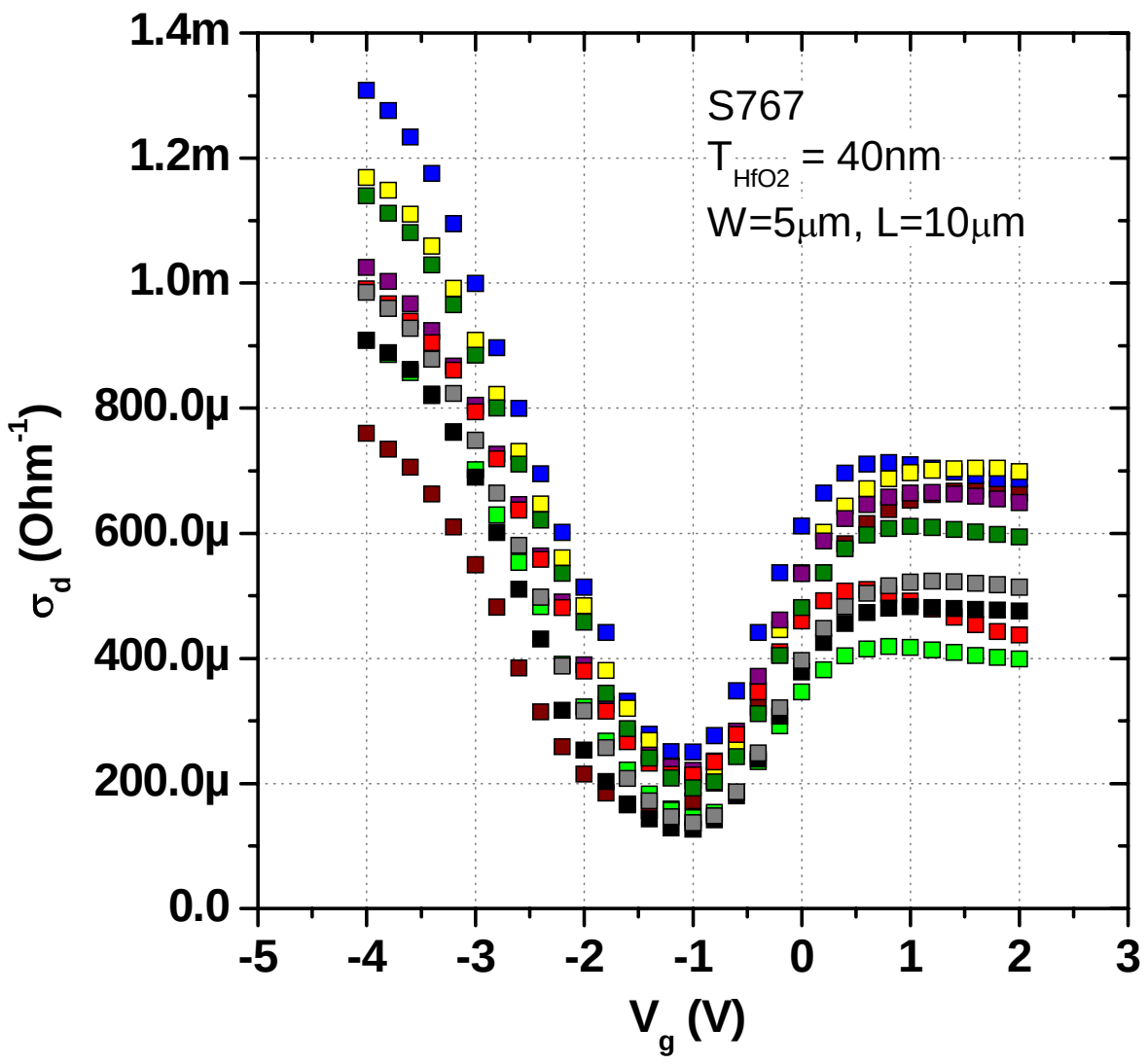

Figure 12: Conductivity characteristics for a set of identically processed devices, for a Si-face sample. On average Si-face devices had much lower minimum conductance and a lower mobility than the C-face samples. Device to device uniformity on the Si-face is much better than on the C-face, with critical device parameters like minimum conductivity and mobility varying only by a factor of 2.

Figure 13:

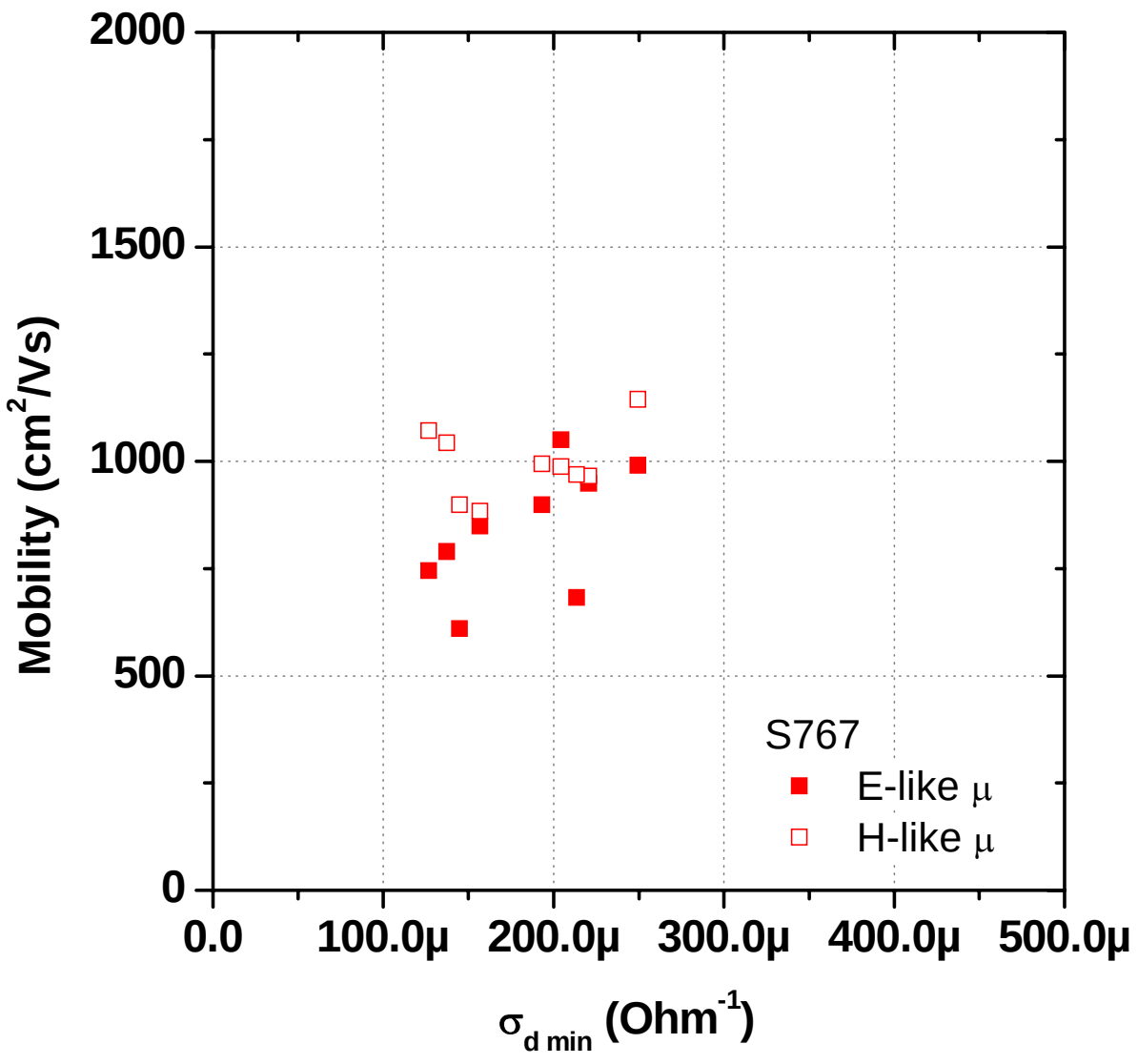

Figure 13: Carrier mobility for Si-face samples plotted vs. minimum conductivity. No strong inter-sample correlation is found.

EOT